\documentstyle[12pt]{article}
\begin{document}
\baselineskip 18pt
\renewcommand{\Re}{{\rm Re \,}}
\renewcommand{\Im}{{\rm Im \,}}
\newcommand{\Tr}{{\rm Tr \,}}
\newcommand{\beq}{\begin{equation}}
\newcommand{\eeq}[1]{\label{#1}\end{equation}}
\newcommand{\bea}{\begin{eqnarray}}
\newcommand{\eea}[1]{\label{#1}\end{eqnarray}}
\newcommand{\dirac}{/\!\!\!\partial}
\newcommand{\Dirac}{/\!\!\!\!D}
\begin{titlepage}
\begin{center}
\hfill hep-th/9711116,  NYU-TH-97/11/01
\vskip .01in \hfill CERN-TH/97-318, UCLA/97/TEP/24
\vskip .4in
{\large\bf Central Extensions of Supersymmetry in Four and Three Dimensions}
\end{center}
\vskip .4in
\begin{center}
{\large Sergio Ferrara}$^{a,b}$~\footnotemark\footnotetext{e-mail:
ferraras@vxcern.cern.ch}
and
{\large Massimo Porrati}$^{c,d}$~\footnotemark\footnotetext{e-mail:
massimo.porrati@nyu.edu}
\vskip .1in
$a$ Theory Division, CERN, CH-1211 Geneva 23, Switzerland
\vskip .05in
$b$ Dept. of Physics UCLA, 405 Hilgard Av. Los Angeles, CA 90024, USA
\vskip .05in
$c$ Department of Physics, NYU, 4 Washington
Pl., New York NY 10003, USA~\footnotemark\footnotetext{Permanent address.}
\vskip .05in
$d$ Rockefeller University, New York NY 10021-6399, USA
\end{center}
\vskip .4in
\begin{center} {\bf ABSTRACT} \end{center}
\begin{quotation}
\noindent
We consider the maximal
central extension of the supertranslation algebra in d=4 and 3, which includes
tensor central charges associated to topological defects such as domain walls
(membranes) and strings. We show that for all $N$-extended superalgebras
these charges are related to nontrivial
configurations on the scalar moduli space.
For $N=2$ theories obtained from compactification on
Calabi-Yau threefolds, we give an explicit realization of the
moduli-dependent charges in terms of wrapped branes.
\end{quotation}
\vfill
November 1997
\end{titlepage}
\section{Introduction}
Central extensions of extended supersymmetry algebras in any dimensions have
been considered in recent years in connection to the dynamics of BPS saturates
p-branes. Generically, for an $N$-extended supersymmetry algebra in $d$
dimensions, the structure of the central extension of the supertranslation
algebra has the form:
\beq
\{ Q^A_\alpha , Q^B_\beta\}=\sum_P Z^{AB}_{\mu_1..\mu_p}(\gamma^{\mu_1..\mu_p}
C)_{\alpha \beta},\;\;\; A,B=1,..,N
\eeq{1}
where, for $p=1$, $Z^{AB}$ must contain a singlet under the automorphism group
of the supercharges (the R-symmetry). This singlet is a linear combination of
the translation generator $P_\mu$ and a singlet vector charge $Z_\mu$. This
central charge can be absorbed in a redefinition of $P_\mu$, but only
when spacetime is uncompactified~\cite{T}. Upon compactification, its spectrum
differs, in general, from the Kaluza Klein spectrum of $P_\mu$.
The total number of central charges,
including $P_\mu$, correctly adds up to the dimension of the symmetric product
of $N$
spinorial representations: $(N\dim s)(N \dim s +1)/2$, where $\dim s$ is
the dimension of an irreducible
spinor representation in space-time dimension $d$.
The existence of tensor charges in the presence of p-branes has been
derived in ref.~\cite{A}; for $N=8$, the algebra in eq.~(\ref{1}) has been
found in~\cite{B}.

The central extension of Poincar\'e supersymmetry limits the central extension
to Lorentz scalars, i.e. $p=0$. On the other hand, already some years ago it
has been realized that p-branes, i.e. extended objects present in consistent
extensions of supergravity theories, can give rise to nonzero tensor central
charges~\cite{A}.

Tensorial central charges can be divided in two categories: charges that
are sources of of dynamical gauge fields, and charges that have no associated
dynamical gauge field, and correspond to topological defects of the theory.

We will study charges in both categories. Particularly: a) vector charges in 4
dimensions, associated to rank-2 antisymmetric tensors, which are dual
to 4-d scalars; b) rank-2 tensor charges, associated to rank 3
antisymmetric tensors, which are non-dynamical in 4-d.
The latter case has received quite some
attention in recent times in the context of supersymmetric Yang-Mills
theories~\cite{S}, since the associated extended object is a BPS saturated
membrane.

In this paper, we first consider the central extension of the supersymmetry
algebra in dimensions 4 and we consider particular physical
situations in which these new central charges are non-vanishing. For the
particular case of type II Calabi-Yau compactifications, we show that the
extended objects source of the tensorial charges lead to new phenomena in the
hypermultiplet moduli space. These phenomena can be described in a general
setting valid for all $N$-extended theories. We also
discuss central extensions of the supertranslation algebra in 3 dimensions,
and its relation to three-dimensional U-duality.
\section{Central Extension of Supersymmetry Algebras in d=4}
The Haag-Lopuszanski-Sohnius central extension of the super-Poincar\'e algebra
for $N$-extended supersymmetry is~\cite{HL}:
\bea
\{Q_\alpha^A, \bar{Q}_{\dot{\alpha} \, B}\} &=&
\sigma^\mu_{\alpha \dot{\alpha}}P_\mu
\delta^A_B, \nonumber \\
\{Q_\alpha^A, Q_\beta^B\} &=& \epsilon_{\alpha \beta}Z^{[AB]}.
\eea{2}
By counting the spinor components in the left-hand side of these equations, it
is evident that some terms have been omitted in the right-hand side. These
additional terms are central charges of the supertranslation algebra,
rather than of the super-Poincar\'e algebra. The complete algebra has
additional terms, as in equation~(\ref{1}), namely
\bea
\{Q_\alpha^A, \bar{Q}_{\dot{\alpha} \, B}\} &=&
\sigma^\mu_{\alpha \dot{\alpha}}P_\mu
\delta^A_B + \sigma^\mu_{\alpha\dot{\alpha}} Z_{\mu\, B}^A,\;\;
(Z^A_{\mu\, A}=0), \nonumber \\
\{Q_\alpha^A, Q_\beta^B\} &=& \epsilon_{\alpha \beta}Z^{[AB]} +
\sigma_{\alpha\beta}^{\mu\nu}Z^{(AB)}_{\mu\nu}.
\eea{3}
These new terms correspond to string charges in the adjoint
of $SU(N)$, and membrane charges in the twofold symmetric of $SU(N)$.

One can easily see that the total number of bosonic generators in the rhs of
eqs.~(\ref{3}) adds up to $8N^2 + 2N$. This is the same dimension as the lhs,
which is a symmetric $4N\times 4N$ matrix.

The tensor charges in eq.~(\ref{3}) can be easily related to geometrical
quantities that depend on the asymptotic value of massless scalar fields
(moduli). To see this, we need only recall that in any theory with local
supersymmetry, the supersymmetry charge can be expressed in a Gauss-like
form which involves only the gravitino field $(\psi_{A\,\mu\alpha}$,
$\bar{\psi}^A_{\mu\dot{\beta}}$):
\beq
Q^A_\alpha=\int_S dx^\mu \wedge dx^\nu \sigma_{\mu\alpha\dot{\alpha}}
\bar{\psi}^{A\, \nu\dot{\alpha}}, \;\;\;
\bar{Q}_{A\, \dot{\alpha}}= \int_S dx^\mu \wedge dx^\nu
\sigma_{\mu\alpha\dot{\alpha}} \psi^\alpha_{A\, \nu}.
\eeq{gauss}
$S$ is a large surface inside a 3-dimensional space-like hypersurface.
Quite generally, one may write the change of the gravitino under a
supersymmetry transformation as:
\beq
\delta_\epsilon \bar{\psi}^A_{\mu \dot{\alpha}}=
D_\mu \bar{\epsilon}^A_{\dot{\alpha}} +
V^A_{B\, \mu}\bar{\epsilon}^B_{\dot{\alpha}} +
{1\over 2}\sigma_{\mu\alpha\dot{\alpha}}M^{AB}
\epsilon_B^\alpha + ... .
\eeq{susy}
An analogous law holds for $\psi_{A\,\mu\alpha}$.
Here, $D_\mu$ is the standard Lorentz-covariant derivative,
the scalar $M^{AB}$ is the gravitino mass matrix, while $V^A_{B\, \mu}$ is
a composite gauge vector, function of the scalar fields in the theory. Notice
that it can be decomposed into a traceless part, in the adjoint of the
R-symmetry $SU(N)$, plus a singlet.
The commutators of two supersymmetry charges can be found most easily by
varying eq.~(\ref{gauss}) with respect to a supersymmetry transformation,
and using eq.~(\ref{susy}). The result is
\bea
\{Q^A_\alpha , \bar{Q}_{B\,\dot{\alpha}}\}&=&  \int_S  dx^\mu \wedge dx^\nu
\sigma_{\mu\alpha\dot{\alpha}} V^A_{B\, \nu} + ... \nonumber \\
\{Q^A_\alpha, Q^B_\beta\} &=& \int_S  dx^\mu \wedge dx^\nu
\sigma_{\mu\nu\, \alpha\beta}M^{AB} + ...
\eea{alg}
Here the ellipsis stand for the standard terms in the supersymmetry algebra
($P_\mu$ and the scalar central charges $Z^{[AB]}$). These equations give
the explicit relation between tensor charges and scalar moduli.

Eqs.~(\ref{alg}) shows that for generic $N$, the central charge of the
domain wall is given in terms of the
gravitino mass matrix, $M^{AB}$, which belongs to the twofold symmetric
representation of the R-symmetry $SU(N)$, while the central charge of the
strings is generated by the composite gauge connection of the R-symmetry.

It is easy to show that in the presence of a string there can exist a nonzero
vector charge $Z_\mu$. Explicitly, in the presence of an infinite string
(either fundamental or solitonic), one may choose as surface of
integration $S$ a cylinder coaxial with the string. For an infinite string in
uncompactified space, the total vector charge diverges, but the charge
density per unit length, $dZ^A_{\mu\, B}/dl$, is finite.
At any point along the string it reads:
\beq
dZ^A_{i\, B}/dl= \hat{l}_i \oint  dx^j V_{j\, B}^A, \;\;\; i,j=1,2,3.
\eeq{loop}
Here $\hat{l}$ is a unit vector tangent to the string and the integral is
taken around any loop orthogonal to $\hat{l}$ and encircling the string once.

Notice that eq.~(\ref{loop}) gives not only the vector charge in the adjoint
of $SU(N)$, but also the singlet vector charge. As we said before, even though
this charge is not algebraically independent from $P_\mu$, and it is
indistinguishable from it in uncompactified space, it becomes an independent
charge upon compactification on non-simply connected manifolds.

Similarly, using eqs.~(\ref{alg}), we can find that the charge
$Z_{\mu\nu}^{(AB)}$ is nonzero in the presence of a domain wall. Consider for
simplicity a flat domain wall located at $z=const$ in Cartesian coordinates.
Then, the second of eqs.~(\ref{alg}) gives us the following expression for
the tensor-charge density (i.e. the charge per unit area of the membrane):
\beq Z_{12}^{(AB)}= [M^{AB}(z=+\infty)-M^{AB}(z=-\infty)], \;\;\;
Z^{(AB)}_{ij}=0\; {\rm otherwise}.
\eeq{memb}

In $N=1$ (rigid) supersymmetry, there exist examples of both BPS
saturated strings (the $N=1$ Abelian Higgs model) and BPS saturated
membranes~\cite{S}. In local supersymmetry, the tensor charges associated to
these objects can be expressed in a model-independent way by using the
superconformal tensor calculus (see for instance~\cite{pvn}),
which allows us to write the off-shell gravitino transformation law as:
\beq
\delta_\epsilon \bar{\psi}_{\mu \dot{\alpha}}=
D_\mu \bar{\epsilon}_{\dot{\alpha}} +
{3\over 4} i A_\mu\bar{\epsilon}_{\dot{\alpha}} -{3\over4}
i\theta(\Phi,\epsilon) \bar{\psi}_{\mu \dot{\alpha}}
-\sigma_{\mu\alpha\dot{\alpha}}\eta^\alpha(\Phi,\epsilon).
\eeq{aux}
In this universal equation, $\epsilon$, $\eta$ are the fermionic generators
of the superconformal group; $A_\mu$ and $\theta$ are, respectively, the
gauge field and the generator of the local R-symmetry. 
The parameters $\eta$ and $\theta$ are not independent,
rather, in any specific model, their dependence on $\epsilon$
and the fields of the theory, $\Phi$,
is fixed by choosing a conformal gauge in which the Einstein action
and the gravitino kinetic term are canonically normalized~\cite{ku}.
The gauge field $A_\mu$ is non-dynamical; 
by solving its algebraic equations of motion 
it can be expressed as a function of the $\Phi$ and their covariant 
derivatives.
Eqs.~(\ref{gauss},\ref{aux}) give a universal formula for the
charge density of strings and domain walls. For domain walls the
formula reads (cfr. eq.~(\ref{memb})):
\beq
Z_{12}= {1\over 2}[\epsilon_{\alpha\beta}{\delta\eta^\alpha\over \delta
\epsilon_\beta}(z=+\infty)-
\epsilon_{\alpha\beta}{\delta\eta^\alpha\over \delta
\epsilon_\beta}(z=-\infty)], \;\;\;
Z_{ij}=0\; {\rm otherwise}.
\eeq{domw}
This formula applies in particular to the the case in which the domain wall is
due to strong-coupling phenomena or fermion condensates~\cite{S}. It holds
also when the supergravity Lagrangian contains higher curvature terms, as the
ones due to string or M-theory corrections.

In $N=2$ supersymmetry, the membrane charges and the algebraically
independent string charges form a triplet of $SU(2)$. The string charges in
the triplet of $SU(2)$ are given by the Wilson loop $\oint  dx^j [
V_{j\, B}^A- (1/2)\delta^A_BV_{j\, C}^C]$. They do not depend at all on the
vector multiplets, since only hypermultiplets contribute to the  composite
gauge connection of the $SU(2)$ R-symmetry~\cite{dW}.
The string charge singlet of
$SU(2)$ instead, is a function of vector multiplets only, since hypermultiplets
do not appear in the $N=2$ formula for $V_{\mu\, C}^C$~\cite{dW}.

The membrane central charge depends on
both hypermultiplet and vector-multiplet moduli, since the formula for the
gravitino mass matrix reads~\cite{dW,F,F2}
\beq
M^{AB}=L^\Lambda P^{AB}_\Lambda,
\eeq{grav}
and $L^\Lambda$ is a covariantly holomorphic section of the $U(1)$ bundle
associated to the vector moduli. This is in agreement with (in fact, it is
required by) the centrally extended algebra in eqs.~(\ref{3}). Indeed, since
$Z^{(AB)}_{\mu\nu}$ enters in the $\{Q^A_\alpha, Q^B_\beta\}$ commutators, it
must carry the same $U(1)$ weight as the scalar charge. This is indeed the
case, since~\cite{F2}
\beq
Z^{[AB]}=\epsilon^{AB}Z, \;\;\; Z= L^\Lambda q_\Lambda - F_\Lambda p^\Lambda.
\eeq{weight}
The string charge $Z^A_{\mu\, B}$, instead, is neutral under $U(1)$, since it
carries the same $U(1)$ as the translations $P_\mu$.

An explicit realization of extended objects with moduli-dependent charge
densities can be obtained in type II superstrings compactified on Calabi-Yau
threefold. The study of these compactifications also provides a check of
the properties discovered above, namely that: a) the charge density of triplet
strings only depends on hypermultiplets, b) the charge density of singlet
strings only depends on vector multiplets, c) the charge density of domain
walls depends on both vector and hypermultiplet moduli.

Let us introduce the following notation: $\mu_1$ is the string charge density,
$\mu_2$ is the domain wall charge density; $J$ is the K\"ahler form of the
CY space, $\Omega$ is the CY holomorphic 3-form, and $a_k$ are the homology
$k$-cycles of the CY space. The 4-d string dilaton, $\hat\phi$, is a 
hypermultiplet both in type IIA and type IIB theory.
It is related to the 10-d dilaton, $\phi$, by
\beq
\hat\phi=\phi + K_J/2,\;\;\; K_J\equiv -\log (\int_{a_6} J\wedge J\wedge J).
\eeq{dil4}
We consider first type IIB superstrings. In this case, the K\"ahler form is
a function of the hypermultiplet moduli only, while $\Omega$ depends only
on the vector moduli.
The 4-D strings present in this theory are:
\begin{enumerate}
\item The fundamental 10-d string. In the 4-d Einstein frame its charge
density is
\beq
\mu_1 = \exp(2\hat\phi).
\eeq{fund}
\item The D-string. Its charge density is
\beq
\mu_1=\exp(\hat\phi)\exp(K_J/2).
\eeq{dstring}
\item The D3 brane wrapped on a 2-cycle. Here:
\beq
\mu_1=\exp(\hat\phi)\exp(K_J/2)\int_{a_2} J.
\eeq{d3}
\item The D5 brane wrapped on a 4-cycle:
\beq
\mu_1=\exp(\hat\phi)\exp(K_J/2)\int_{a_4} J\wedge J.
\eeq{d5}
\item The D7 brane wrapped on the CY threefold: 
\beq
\mu_1=\exp(\hat\phi)\exp(-K_J/2).
\eeq{d7}
\item The NS5 brane wrapped on a 4-cycle.
\beq
\mu_1=\exp(K_J)\int_{a_4}J \wedge J.
\eeq{ns5}
\end{enumerate}
All the charge densities listed above are functions of hypermultiplets only;
thus all the strings of type IIa on CY are triplets.

To get a domain wall in 4-d, on the other hand,
one must wrap a D5 or NS5 brane on a 3-cycle.
The charge densities now read
\beq
\mu_2 = \exp (\hat\phi)\exp(K_J)V_3, \;{\rm for\; NS5},\;\;\;
\mu_2 = \exp (2\hat\phi)\exp(K_J/2)V_3, \;{\rm for\; D5}.
\eeq{DNS5}
The volume of the minimal-volume 3-cycle is~\cite{SBB}
\beq
V_3=\exp(K_\Omega/2-K_J/2)|\int_{a_3} \Omega|, \;\;\; K_\Omega= -\log (i\int
\bar{\Omega} \wedge \Omega).
\eeq{becker}
Since $V_3$ depends on vector moduli, $\mu_2$ is a function of both vectors
and hypermultiplets, as predicted by the general SUSY algebra in eq.~(\ref{3}).

It is interesting to compare the above formulae with the general formula 
eq.~(\ref{loop}), in which the charges are expressed by integrating the
$SU(2)$ connection of the quaternionic manifold.
As examples, let us consider the fundamental string, the D-string, and the NS 
5-brane wrapped on a 4-cycle.
For this purpose, we may use the general formulae~\cite{FS} of quaternionic
manifolds obtained by c-map from special manifolds~\cite{CFG}. By writing the 
$SU(2)$ connection in terms of sigma matrices,
\beq
V^A_B=V^I\sigma^A_{I\,B},
\eeq{s1}
the differentials from the NS-NS sector give a contribution to $V^3$, 
while the R-R differentials contribute to $V^1,V^2$. For the fundamental and
D-strings in the universal hypermultiplet we get, respectively:
\beq
V^3\sim(S+\bar{S})^{-1}da,\;\;\; 
V^1\sim (S+\bar{S})^{-1/2}\exp({K_J/2})dC.
\eeq{s2}
Here $a$ is the space-time axion and $S\equiv ia + \exp(-2\hat{\phi})$.
Integrating these formulas we reproduce eqs.~(\ref{fund},\ref{dstring}).
Also, for the NS 5-brane we obtain
\beq
V^3 \sim Q,
\eeq{s3}
where $Q$ is the connection one-form of the $U(1)$ Hodge bundle of the special
geometry associated by c-map to the quaternionic manifold. In the
case at hand, in special coordinates,
\beq
Q\sim \exp(K_J) d_{ABC}\Re Z^A \Re Z^B d \Im Z^C.
\eeq{s4}
Integrating $Q$ we reproduce eq.~(\ref{ns5}).

The type IIA string provides us with
another subtle check of our formulae. In type IIA
superstrings, indeed, $J$ is a function of the vector moduli, while $\Omega$
depends on the hypermultiplet moduli only.
The 4-d strings in this case come from:
\begin{enumerate}
\item The fundamental 10-d string, $\mu_1=\exp(2\hat\phi)$.
\item D4 branes wrapped on 3-cycles, with charge density:
\beq 
\mu_1=\exp(\hat\phi)\exp(K_J/2)V_3=\exp(\hat\phi)
\exp(K_\Omega/2)|\int_{a_3} \Omega|.
\eeq{d4a}
\item NS5 branes wrapped on 4-cycles:
\beq
\mu_1=\exp(K_J)\int_{a_4} J\wedge J.
\eeq{NS5a}
\end{enumerate}
Notice that the string obtained from the D4 brane is a triplet of $SU(2)$,
since its charge density eq.~(\ref{d4a}) depends only on hypermultiplets,
but the string obtained from the NS5
brane is a singlet, since its $\mu_1$ depends only on vector moduli.

Finally, the 4-d domain walls of type IIA are:
\begin{enumerate}
\item D2 branes. Their charge density is: $\mu_2=\exp(2\hat\phi)\exp(K_J/2)$
\item D4 branes wrapped on 2-cycles: $\mu_2=\exp(2\hat\phi)\exp(K_J/2)
\int_{a_2}J$.
\item D6 branes wrapped on 4-cycles: $\mu_2=\exp(2\hat\phi)\exp(K_J/2)
\int_{a_4}J\wedge J$.
\item D8 branes wrapped on the CY threefold: 
$\mu_2=\exp(2\hat\phi)\exp(-K_J/2)$.
\item NS5 branes wrapped on 3-cycles: $\mu_2$ is given by the first equation
in formula~(\ref{DNS5}).
\end{enumerate}
As expected, they depend on both hypermultiplets and vector multiplets.
\section{Central Extension in d=3}
The analysis done above for four-dimensional extended supersymmetry can be
repeated in three dimensions. In d=3, the R-symmetry of the $N$ extended
supertranslation algebra is $O(N)$, rather than $SU(N)$, and its maximal
central extension reads
\beq
\{ Q_\alpha^A, Q_\beta^B\}= (\gamma^\mu C)_{\alpha\beta}P_\mu\delta^{AB} +
 (\gamma^\mu C)_{\alpha\beta}Z_\mu^{(AB)} + C_{\alpha\beta}Z^{[AB]}, \;\;
(Z^{(AB)}\delta_{AB}=0).
\eeq{4}
We see again that by adding to the translations a vector charge,
traceless and symmetric under $O(N)$, together with a scalar charge, in the
adjoint of $O(N)$, we find $N(2N+1)$ bosonic charges. This is the dimension of
the lhs of eq.~(\ref{4}).

The adjoint of $O(2N)$ decomposes as follows under $U(N)\subset O(2N)$:
\beq
adj \, O(2N) = adj \, U(N) + [2] + [\bar{2}],
\eeq{5}
where $[2]$ denotes the twofold antisymmetric of $U(N)$. This property shows
that the scalar central charges in 3 dimensions come from the scalar central
charges in four dimensions plus the extra scalar charges coming from the
dimensional reduction of $Z_{\mu\, B}^A$ and $P_\mu$.

The simplest model with a nonzero central charge $Z^{[AB]}$ is the $N=2$
Abelian Higgs model. It is obtained by dimensional reduction of the
4-d Abelian Higgs model, which contains a BPS saturated string. Upon
dimensional reduction, the associated string charge gives rise to a scalar
central charge $Z^{[AB]}=\epsilon^{[AB]}Z$. Physically, this is the charge of
the 3-d vortex generated by the dimensionally reduced BPS string.

In 3 dimensions, the U-duality group of the $N=16$ theory is $E_{8(8)}$.
This strongly suggests that
the 120 scalar charges --in the adjoint of $O(16)$-- together with the
charges of the 128 ``elementary'' massless fields --in the left spinor
representation of $O(16)$-- should complete the 248-dimensional fundamental
representation of $E_{8(8)}$.
\section{Conclusions}
In this paper, we found an explicit universal formula relating the tensor
central charges to the geometry of scalar moduli fields. These formulae are
easily generalized to take into account the possible existence of non-vanishing
fermion condensates. We presented this generalization explicitly for the
(off-shell) $N=1$ superalgebra.
Condensates are expected in strongly interacting
theories. In rigid supersymmetry, indeed, it has been shown explicitly that
they give rise to membrane central charges~\cite{S}.

We must also note that all formulae for string tensions of type IIB on CY 
spaces obtained in this paper receive string quantum corrections, because the 
quaternionic geometry is corrected both perturbatively~\cite{S2} and
non-perturbatively~\cite{SBB,OO,S3,AFMN}. 
However, eq.~(\ref{loop}) is non-perturbative in
nature, and allows to determine the charge of all 4-d strings in terms of the
quantum-corrected quaternionic geometry. This is parallel to the case of zero
branes, where the central-charge formula~\cite{F2} allows to study 
non-perturbative properties of type IIB strings near a conifold 
point~\cite{S3},
or fixed scalars at the black hole horizon~\cite{F3}. 

Finally, we must point out that when the volume of the CY space is finite, all 
moduli are dynamical. The 4-d strings and membranes modify the equations of
motion of these scalars, driving them to some fixed points. An interesting
question, that we would like to address in the future, is whether 
fixed points exist where the string or domain-wall tension is neither zero
nor infinite. These points exist in the analogous case of 4-d zero branes, 
where moduli can flow to fixed point with finite mass~\cite{F3}.
Tree-level formulae as in eqs.~(\ref{fund}-\ref{ns5}) do
not allow for nontrivial fixed points, but the quantum corrected ones may. 
\vskip .2in
\noindent
Acknowledgments
\vskip .1in
\noindent
We would like to thank T. Banks for useful correspondence.
S.F. is supported in part by DOE under grant DE-FGO3-91ER40662, Task C, and by
EEC Science Program SC1*-CI92-0789.
M.P. is supported in part by NSF under grant PHY-9722083.

\end{document}